\documentclass[twocolumn]{aastex631}

\usepackage{physics, bm}
\newcommand{\revise}[1]{{#1}}

\begin{document}

\title{Semi-analytical model for the dynamical evolution of planetary system II: Application to systems formed by a planet formation model}

\author[0000-0001-8477-2523]{Tadahiro Kimura}
\email{tad.kimura624@gmail.com}
\affiliation{UTokyo Organization for Planetary Space Science (UTOPS), University of Tokyo, Hongo, Bunkyo-ku, Tokyo 113-0033, Japan}
\affiliation{Kapteyn Astronomical Institute, University of Groningen Landleven 12, 9747 AD, Groningen, Netherlands}
\affiliation{Division of Science, National Astronomical Observatory of Japan,
Osawa, Mitaka, Tokyo 181-8588, Japan}

\author[0000-0002-5486-7828]{Eiichiro Kokubo}
\affiliation{Division of Science, National Astronomical Observatory of Japan,
Osawa, Mitaka, Tokyo 181-8588, Japan}
\affiliation{Center for Computational Astrophysics, National Astronomical Observatory of Japan, 
Osawa, Mitaka, Tokyo 181-8588, Japan}
\affiliation{Department of Astronomy, University of Tokyo,
Hongo, Bunkyo-ku, Tokyo 113-0033, Japan}
\affiliation{Graduate Institute for Advanced Studies, SOKENDAI, 2-21-1 Osawa, Mitaka, Tokyo 181-8588, Japan}

\author[0000-0002-2383-1216]{Yuji Matsumoto}
\affiliation{Center for Computational Astrophysics, National Astronomical Observatory of Japan, 
Osawa, Mitaka, Tokyo 181-8588, Japan}

\author[0000-0002-1013-2811]{Christoph Mordasini}
\affiliation{Division of Space Research and Planetary Sciences, Physics Institute, University of Bern, Gesellschaftsstrasse 6, 3012 Bern, Switzerland}

\author[0000-0002-5658-5971]{Masahiro Ikoma}
\affiliation{Division of Science, National Astronomical Observatory of Japan,
Osawa, Mitaka, Tokyo 181-8588, Japan}
\affiliation{Graduate Institute for Advanced Studies, SOKENDAI, 2-21-1 Osawa, Mitaka, Tokyo 181-8588, Japan}
\affiliation{Department of Earth and Planetary Science, University of Tokyo, Hongo, Bunkyo-ku, Tokyo 113-0033, Japan}



\begin{abstract}
The standard formation model of close-in low-mass planets involves efficient inward migration followed by growth through giant impacts after the protoplanetary gas disk disperses. 
While detailed $N$-body simulations have enhanced our understanding, their high computational cost limits statistical comparisons with observations.
In our previous work, we introduced a semi-analytical model to track the dynamical evolution of multiple planets through gravitational scattering and giant impacts \revise{after the gas disk dispersal}. Although this model successfully reproduced $N$-body simulation results under various initial conditions, \revise{our} validation was still limited to cases with compact, \revise{equally-spaced} 
planetary systems.
In this paper, we improve our model \revise{to handle more diverse planetary systems characterized by broader variations in planetary masses, semi-major axes, and orbital separations}
and validate it against recent planet population synthesis results. Our enhanced model accurately reproduces the mass distribution and orbital architectures of the final planetary systems.
Thus, we confirm that the model can predict the outcomes of post-gas disk dynamical evolution \revise{across a wide range of planetary system architectures},
which is crucial for reducing the computational cost of planet formation simulations.
\end{abstract}



\section{Introduction} \label{sec:intro}
One major type of detected exoplanets are those with radii of 1--4$R_\oplus$ and masses 1--$20M_\oplus$ at short orbital periods $<$~100~days~\citep[e.g.,][]{Mayor+2011,Howard+2012,Petigura+2017}, often called super-Earths or sub-Neptunes. Planetary systems with multiple super-Earths and sub-Neptunes are also common~\citep[e.g.,][]{Howard2013,Zhu+2018}. 
Prevailing formation models suggest that these planets undergo efficient inward migration~\citep[e.g.,][]{Papaloizou+Larwood2000,Brunini+Cionco2005,Terquem+Papaloizou2007,Mordasini+2009,Ida+Lin2010,Coleman+Nelson2014,Cossou+2014, Huang+Ormel2022}, followed by growth through giant impacts after the protoplanetary disk disperses~\citep[e.g.,][]{Ogihara+Ida2009,McNeil+Nelson2010,Izidoro+2017}.
Detailed $N$-body simulations of the giant impact phase following gas disk dispersal have advanced our understanding of the formation of these planets~\citep[e.g.,][]{Hansen+Murray2013,Matsumoto+2021,Goldberg+Batygin2022}.
However, the high computational cost of these simulations limits the number of calculations for statistical comparison with observations \citep{Emsenhuber+2021b}.

Some previous studies have developed more computationally efficient models for the dynamical evolution of multi-planetary systems using analytical plus Monte Carlo approach~\citep{Ida+Lin2010,Ida+2013,Kimura+2022} or Machine Learning techniques~\citep{Lammers+2024}.
Although these models reproduce $N$-body simulation results under specific initial conditions, developing models applicable to various planetary systems remains a key challenge.

In our previous study (hereafter Paper I), we developed a new semi-analytical model to follow the dynamical evolution of multiple planets through giant impacts and gravitational scatterings after gas disk dispersal.
We showed that the model could reproduce the results of direct $N$-body simulations under various initial conditions, such as central stellar masses, planetary masses, and orbital radii.
However, validation was limited to close-in ($\lesssim 1$~au), compact, equally spaced, similar-mass planetary systems.
Recent planet formation models show diverse planetary distributions depending on initial conditions~\citep{Emsenhuber+2021b}. Planets are spread over a wide range of orbital radii, with large variations in masses and orbital spacings, and some are in mean motion resonances. Thus, it is important to test the applicability of our semi-analytical model to more general planetary systems.

In this paper, we improve our model for general planetary systems and validate it by comparing it with recent results of the planet population synthesis~\citep[][]{Emsenhuber+2021b,Emsenhuber+2023}. 
In \S~\ref{sec:method}, we detail the semi-analytical model, focusing on updates from Paper I. The data of the $N$-body-based planet formation model and the comparison method are given in \S~\ref{sec:Nbody-data}, and the comparison results are presented in \S~\ref{sec:results}. We discuss some model limitations in \S~\ref{sec:discussion}, and finally, in \S~\ref{sec:conclusion}, we summarize our results and discuss future perspectives.

\section{Dynamical evolution model} \label{sec:method}

In this study, we improve the semi-analytical model that we developed in Paper I
\revise{to} apply it to diverse planetary systems.
\revise{
The original model in Paper I incorporates secular evolution of planetary eccentricities, predicts the timing of orbital crossing events, and determines whether a giant impact or close scattering occurs thereafter. The model then evaluates the resultant orbital elements after each event. By iteratively applying these steps, the model simulates the post-gas-disk long-term dynamical evolution and final configuration of planetary systems.
}
\revise{However, the model was designed for planetary systems that are compactly packed 
and do not undergo planet ejection.
To improve its applicability to more diverse planetary architectures, here we introduce the following extensions:}
(1) grouping planets in a system to evaluate the number of planets contributing to the onset of the orbital instability, 
and (2) \revise{incorporating the possibility of planet ejection} 
through close scattering events.
We describe each model in the following.
In this paper, we denote the mass, radius, semi-major axis, and eccentricity of planet $i$ as $M_i, R_i, a_i$,and $e_i$, respectively.

\subsection{Grouping of planets}

The semi-analytical model predicts the onset time of orbital instability for three neighboring planets (orbital crossing timescale $\tau_{\rm cross}$) using the formula by \cite{Petit+2020}. Since this formula is intended for three-planet systems, we introduce a factor $K$ for the density of three-body resonances to extend its applicability to general $N$-body systems. In Paper I, we evaluated $K$ as
\begin{equation}
    K = \min\qty( 3, 0.5(N-3) + 1),
\end{equation}
where $N$ is the number of planets in the system. This assumes that up to two neighboring planets, both inside and outside the triplet, can affect $\tau_{\rm cross}$, which is valid for compact systems. However, in general systems, planets may be widely separated, and neighboring planets of a triplet do not necessarily affect its stability. Thus, only planets with relatively narrow spacing should be counted in $N$.

To address this, we divide the planets in a system into groups. Recently, Kokubo et al. (submitted) found that the orbital architecture of multi-planet systems formed by giant impacts can be scaled with
\begin{equation}
    r_{{\rm K}, ij} = e_{{\rm esc}, ij} a_{ij},
\end{equation}
where $a_{ij} = (a_i+a_j)/2$, and $e_{{\rm esc},ij}$ is the escape eccentricity defined as \citep[e.g.,][]{Safronov1969, Kokubo+Ida2002}
\begin{equation}
    e_{{\rm esc},ij} = \frac{v_{\rm esc}}{v_{\rm K}}
= \frac{\sqrt{2G(M_i+M_j)/(R_i+R_j)}}{\sqrt{GM_*/a_{ij}}},
\label{eq:e_esc}
\end{equation}
with $v_{\rm esc}$ and $v_{\rm K}$ being the two-body surface escape velocity and the Kepler velocity, respectively. Kokubo et al. found that the orbital separation of planets after collisional evolution is characterized by
$b_{{\rm K},ij} = (a_j-a_i)/r_{{\rm K},ij} \sim$2.0--2.5, and this typical value is almost independent of the initial conditions. Thus, we set a critical value of $b_{{\rm K},ij}$ as $b_{\rm K,crit} = 2.5$, and divide the planets into groups where the orbital spacing is $b_{{\rm K},ij} > b_{\rm K,crit}$. For example, if planets 1, 2, 3, 4, and 5 have $b_{{\rm K},12}, b_{{\rm K},23}, b_{{\rm K},45} < b_{\rm K,crit}$, and $b_{{\rm K},34} > b_{\rm K,crit}$, they are grouped into (1, 2, 3) and (4, 5). The number of planets in the same group is used as $N$ when calculating $\tau_{\rm cross}$ of a triplet. We confirmed that the choice of $b_{\rm K,crit}$ between 2--3 does not affect our results.

\subsection{Ejection of planets}
In some cases, especially in the outer region, the eccentricity after a close scattering event can exceed unity, leading to the ejection of the planet from the system. This ejection event is calculated similarly to \cite{Ida+2013}. Firstly, the planet with $e \ge 1$ (if both interacting planets have $e \ge 1$, the one with the larger eccentricity is selected) is removed from the system. The semi-major axis of the remaining planet is then calculated assuming energy conservation as follows~\citep[e.g.,][]{Rasio+Ford1996,Marzari+Weidenschilling2002,Ford+Rasio2008}:
\begin{equation}
    \frac{M_{\rm rem}}{a_{\rm rem}}
    = \frac{M_{\rm rem}}{a_{{\rm rem},0}}
    +\frac{M_{\rm ejc}}{a_{{\rm ejc},0}},
\end{equation}
where the subscripts ``rem'' and ``ejc'' indicate the remaining and ejected planet, respectively, and the subscript 0 indicates the value just before the scattering event. 
We assume that the apocenter distance of the remaining planet is equal to its semi-major axis before the scattering, allowing us to evaluate the eccentricity after the scattering as
\begin{equation}
    e_{\rm rem} = \frac{a_{{\rm rem},0}}{a_{\rm rem}} - 1.
\end{equation}

\section{Validation method of our model}
\label{sec:Nbody-data}
\begin{figure}
    \centering
    \includegraphics[width=\linewidth]{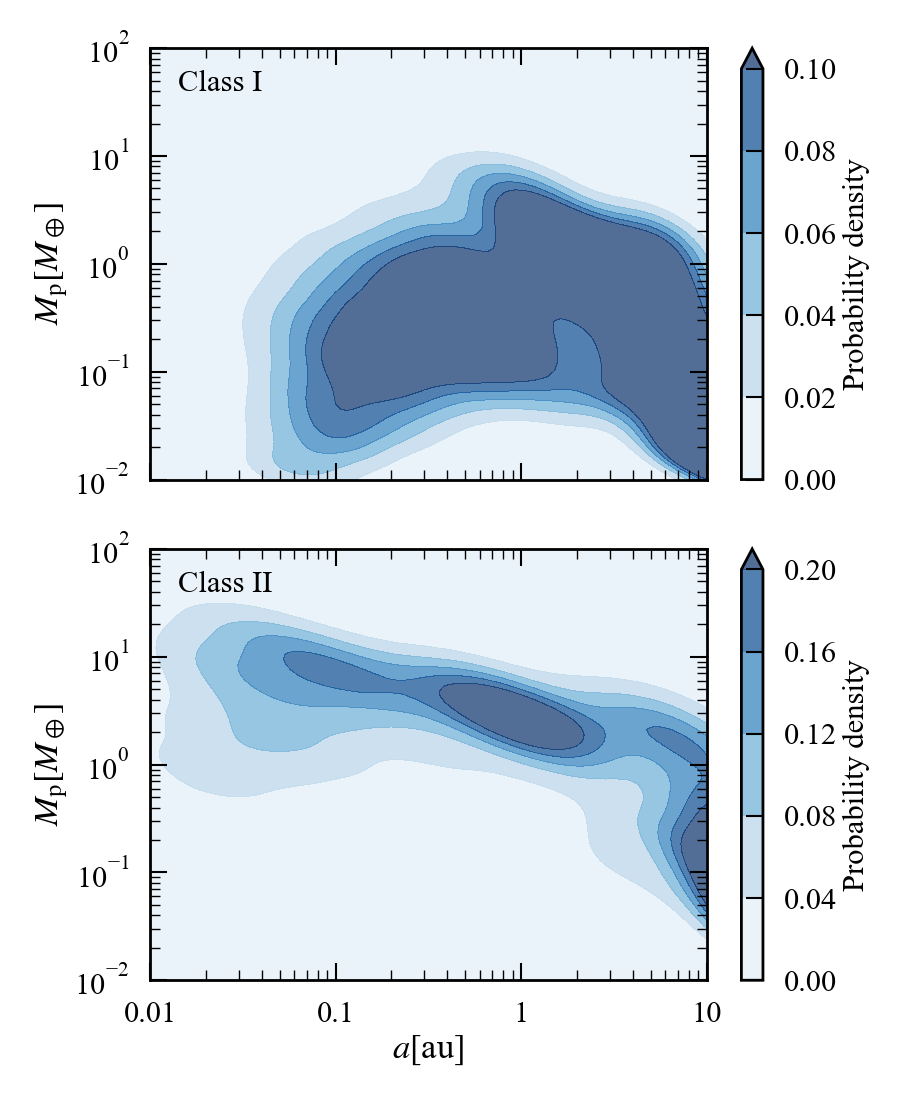}
    \caption{Probability density distribution of the mass and semi-major axis of planets at the time of gas disk dispersal for Class I (compositionally ordered Earth-like and icy planet systems) and Class II (migrated icy sub-Neptune systems) populations in the NGPPS model, which are used as the initial conditions in our calculations. See \S~\ref{sec:numerical_setting} for details.
    }
    \label{fig:Mp-a_init}
\end{figure}

To demonstrate the applicability of our semi-analytical model to various planetary systems, we compare the model outcomes with the results of the New Generation Planet Population Synthesis ~\citep[NGPPS;][]{Emsenhuber+2021b,Emsenhuber+2023}. These populations were generated with the Generation III Bern Model of Planet Formation and Evolution \citep{Alibert+2005,Mordasini+2009,Mordasini+2012,Alibert+2013,Emsenhuber+2021a}. This global model uses direct $N$-body integrations to follow planet-planet gravitational interactions. We use the nominal population data (\verb|NG76longshot|) from \cite{Emsenhuber+2023}. The numerical settings, procedures, and assumptions in our calculations are described below.

\subsection{Numerical Settings}\label{sec:numerical_setting}
\revise{Since our model is designed to track the dynamical evolution of planetary systems after the gas disk dispersal, we focus on the post-disk phase in the comparison with NGPPS results as well. Specifically, for each planetary system computed in NGPPS, we take the system configurations (i.e. planetary masses, radii, semi-major axes, and eccentricities) at the time of gas disk dispersal (typically at $\sim 3$~Myrs) as the initial condition. We then use our model to simulate the subsequent evolution of each system and compare the resulting final state with that obtained in NGPPS. }
The integration time of our model is set to 100 Myrs, matching the duration of the $N$-body calculations in the NGPPS calculations \citep{Emsenhuber+2023}. The central stellar mass is fixed at $1M_\odot$.

Since the NGPPS data include various types of planetary system, we first classify the systems into four classes, following \cite{Emsenhuber+2023}:
I) Systems composed of low-mass planets (of $\lesssim 10 M_\oplus$) with limited migration, most of which grow approximately in situ. These systems are compositionally ordered with rocky Earth-like planets inside and icy planet outside. These systems are shaped mainly by the final phase of giant impacts;
II) those composed of low-mass icy planets (of $\lesssim 30 M_\oplus$), most of which undergo long-distance migration from beyond the water iceline to close to the host star. These systems are shaped by migration in resonant convoys, followed in part by a breaking of the resonances;
III) those hosting both inner low-mass planets and outer giant planets;
IV) those with giant planets which underwent strong dynamical interactions. Almost no inner low-mass planets exist in these systems.
In this study we focus on Classes~I \revise{(576 systems)} and II \revise{(221 systems)} for comparisons, as our semi-analytical model is designed for planetary systems without giant planets. The initial planet populations of Classes I and II (at the time of gas disk dispersal) are shown in Fig.~\ref{fig:Mp-a_init}. The existence of these two classes with distinct formation pathways and resulting compositions might be responsible \citep{Burn+2024} for the observed radius valley \citep{Fulton+2017,VanEylen+2018}.
The probability density is calculated using the 2D Gaussian kernel density estimation (KDE), with the bandwidth determined by the standard Scott rule~\citep{Scott1992}. 

\begin{figure*}
    \centering
    \includegraphics[width=\linewidth]{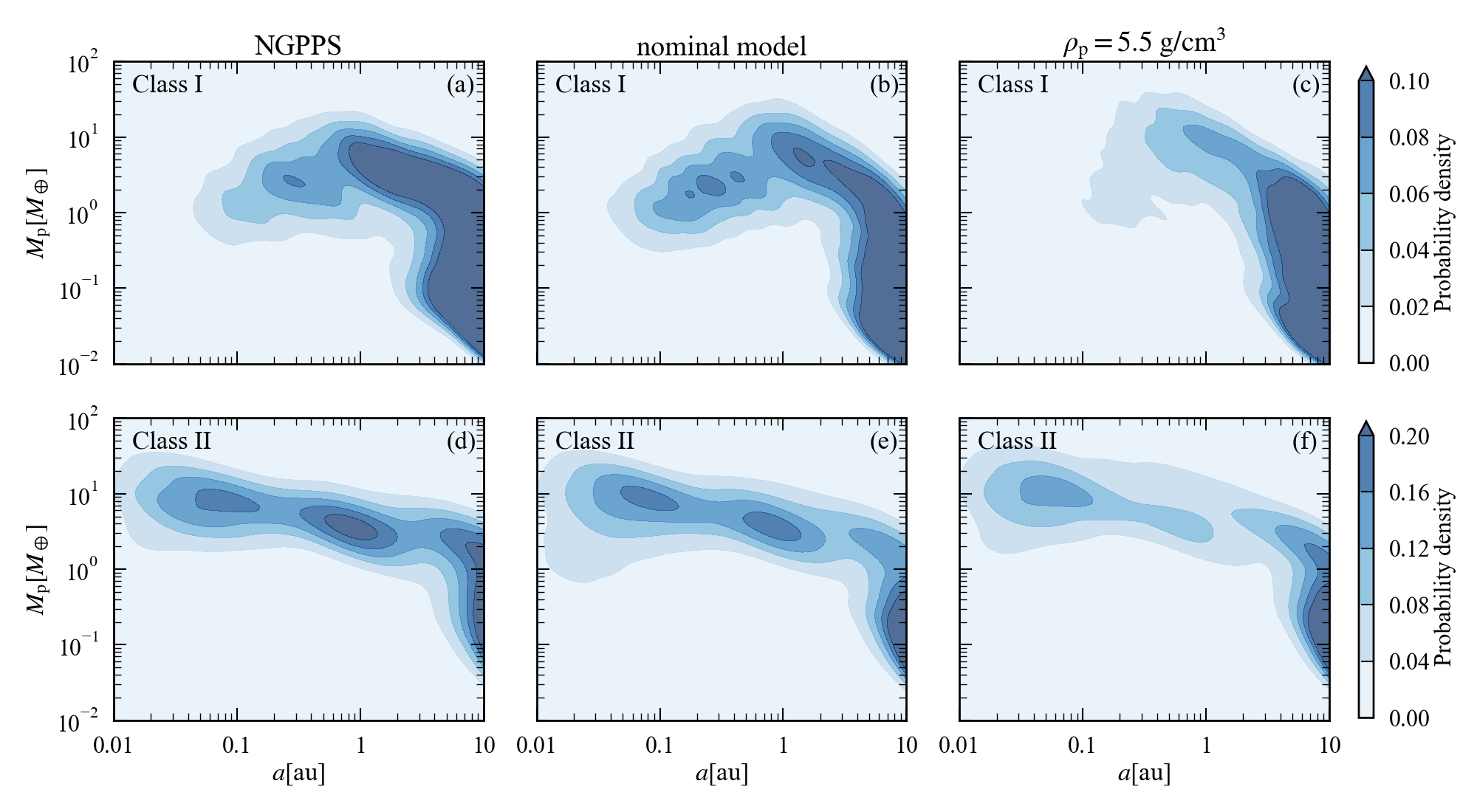}
    \caption{Probability density distributions of the final mass and semi-major axis of simulated planets from the NGPPS model (panels (a) and (d)) and our model (panels (b), (c), (e) and (f)). Panels (a)--(c) show the cases for Class I, while panels (d)--(f) show those for Class II.
    The middle ((b) and (e)) and right ((c) and (f)) columns present results for the nominal model and the model assuming a constant bulk density of $\rho_{\rm p} = 5.5~{\rm g/cm^3}$, respectively (see \S~\ref{sec:planet_radius} for details on panels (c) and (f)).
    }
    \label{fig:Mp-a_comparison}
\end{figure*}
\subsection{Assumptions}
While our model focuses on giant collisions and close encounters between 
planets, the NGPPS includes many other processes, such as envelope thermal evolution and escape, growth via planetesimal accretion, and tidal evolution of orbits.
To conduct our model calculations 
as closely as possible to the conditions of the NGPPS calculations, we make the following assumptions: 
\begin{enumerate}
    \item Planetary radii \\ 
        In the NGPPS calculations, a giant collision is identified when the physical separation between the centres of two planets is smaller than the sum of their planetary radii, which include the thickness of their envelopes. 
        Immediately after the gas disk dispersal, planets are significantly inflated due to high accretion energy, with more than half having bulk densities of less than $0.1~{\rm g/cm^3}$. The planetary radii then rapidly shrink on a $\sim$ Myr timescale due to radiative cooling and envelope escape. 
        To account for envelope thickness, we assume that the planets retain their initial radii (i.e., the radii at the time of disk dispersal) until $t=1$~Myr, which is close to the typical envelope contraction timescale (Kelvin-Helmholtz timescale) for these planets, or until the first collision event. After that, we calculate the planetary radius, $R_{\rm p}$, using the fitting formula by~\cite{Lopez+Fortney2014};
        \begin{equation}
            R_{\rm p} = R_{\rm core} + \Delta R_{\rm env}
        \end{equation}
        where $R_{\rm core}$ is the solid core radius given by
        \begin{equation}
            R_{\rm core} = \qty(\frac{M_{\rm core}}{M_\oplus})^{0.25}R_\oplus
        \end{equation}
        with $M_{\rm core}$ being the solid core mass, and $\Delta R_{\rm env}$ is the envelope thickness given by
        \begin{align}
            \Delta R_{\rm env}
            &= 2.06 \qty(\frac{M_{\rm p}}{M_\oplus})^{-0.21}
            \qty(\frac{f_{\rm env}}{5\%})^{0.59} \notag \\
            & \qquad \times \qty(\frac{F_{\rm p}}{F_\oplus})^{0.044}
            \qty(\frac{t}{5~{\rm Gyr}})^{-0.11} R_\oplus.
        \end{align}
        Here $f_{\rm env}$ is the envelope mass fraction relative to the core mass, and $F_{\rm p}$ is the incident flux, calculated from the present-day solar luminosity $L_\odot$ and the semi-major axis.
        Note that we ignore the envelope mass loss due to photo-evaporation and giant collisions.
        The effects of inflated radii will be discussed later.
    \item Ignoring planetesimal accretion \\
        In the NGPPS simulations, some planets continue to grow through planetesimal accretion 
        even after disk dispersal, 
        especially in outer regions ($\gtrsim$ a few au). In our calculations, we ignore planetesimal accretion, assuming that 
        planets grow only by mutual collisions. Although this can affect the final mass of outer planets and, through secular perturbations, the evolution of inner planets, we have confirmed that the impact on the final distribution is small. 
\end{enumerate}

\subsection{Statistics of individual system to be compared}
As in Paper I, we focus on the following statistical features of the final distribution of planets for comparisons:
i) the final number of planets, ii) the mean orbital separation and eccentricity, both normalised by the mutual Hill radius, namely, 
\begin{align}
    \bar{b}_{\rm H} &= \frac{1}{N_{\rm p}-1}\sum_{i}^{N_{\rm p}-1} \frac{a_{i+1}-a_i}{r_{{\rm H},ij}},\\
    \bar{e}_{\rm H} &= \frac{1}{N_{\rm p}-1}\sum_{i}^{N_{\rm p}-1} \frac{e_i a_i+e_{i+1}a_{i+1}}{2r_{{\rm H},ij}},
\end{align}
where $r_{{\rm H},ij}$ is the mutual Hill radius for planets $i$ and $j$, and iii) the normalized standard deviation of planet mass for each system:
\begin{equation}
    \frac{\sigma_M}{\bar{M}} =\frac{1}{\bar{M}} \sqrt{\frac{1}{N_{\rm p}}\sum_i^{N_{\rm p}}(M_i-\bar{M})^2}
\end{equation}
Here $\bar{m}$ represents the average value of the quantity $m$ in each system, while $\langle m \rangle$ denotes the average value calculated across all systems.

\section{Results} \label{sec:results}
\begin{figure}
    \centering
    \includegraphics[width=0.9\linewidth]{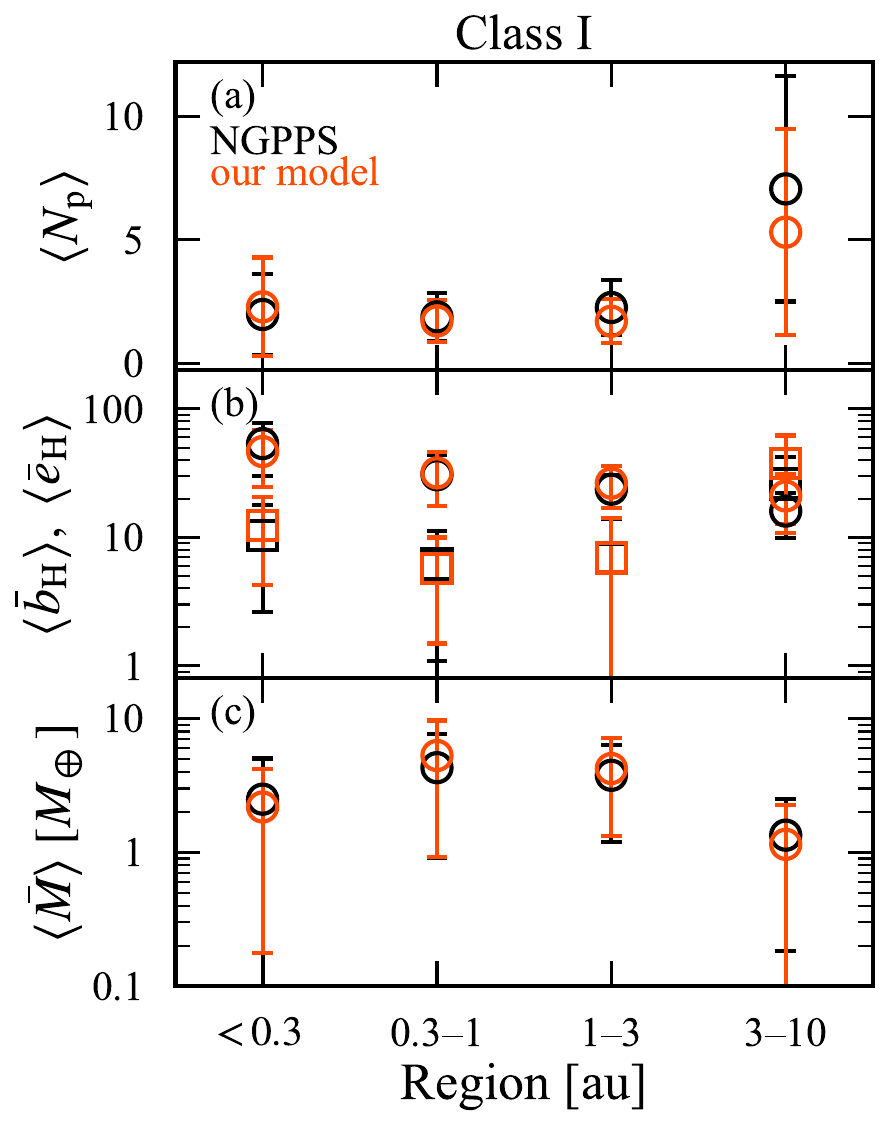}
    \caption{Statistical properties of planets within each orbital radius region for the Class I population.
    The red and black symbols with error bars represent results from our semi-analytical model and the NGPPS model, respectively.
    Each panel shows the mean (symbols) and standard deviation (error bars) of (a) the final number of planets $N_{\rm p}$, (b) the mean orbital separation $\bar{b}_{\rm H}$ (circle) and mean eccentricity $\bar{e}_{\rm H}$ (square), both scaled by the mutual Hill radius, and (c) \revise{
    the average planetary mass $\bar{M}$.
    }
    }
    \label{fig:all_stats_classI}
\end{figure}
\begin{figure}
    \centering
    \includegraphics[width=0.9\linewidth]{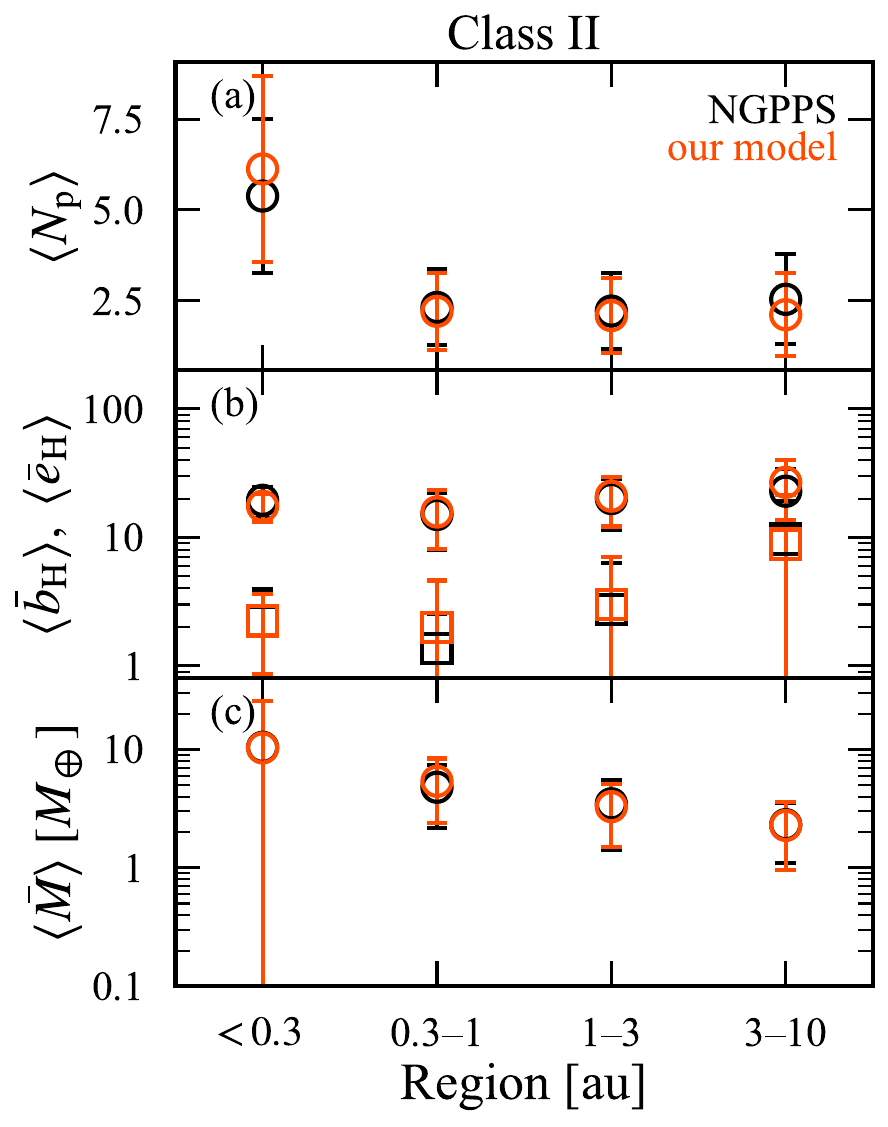}
    \caption{Same as Fig.~\ref{fig:all_stats_classI}, but for the Class II population.
    }
    \label{fig:all_stats_classII}
\end{figure}
\subsection{Final distribution of planets}

We present a comparison between the outcomes from our semi-analytical model and those from the NGPPS model. Figure~\ref{fig:Mp-a_comparison} illustrates the probability density distribution of the mass and semi-major axis of simulated planets in the final state (i.e., at 100~Myrs). 
Here we focus on the comparison between our nominal model (panels~(b) and (e)) and the NGPPS model (panels~(a) and (d)).

For Class I systems, both models yield significant changes in planetary distribution from the initial state (see Fig.~\ref{fig:Mp-a_init}). In the NGPPS calculations (panel~(a)), many of the initially present sub-Earth-mass planets have grown through frequent mutual collisions, leading to a reduction in the number of planets, an increase of super-Earths, and a smaller overall mass variation. This trend is well reproduced in our semi-analytical model, as shown in panel~(b). 
These changes can be interpreted as follows: 
Since most planets in Class~I have not undergone significant migration, planetary mass is mostly determined by the local isolation mass, which increases with orbital radius.
Consequently, the orbits of less massive inner planets are easily disturbed by massive outer planets through secular perturbations, leading to efficient collisional growth, especially in the inner region.
Note that the lack of super-Earth-mass planets in the outer region ($\gtrsim$~3~au) in our model is due to the exclusion of mass growth by planetesimal accretion.


Additionally, we have found that 
our semi-analytical model tends to yield slightly larger masses in the distant region ($\sim$1--3~au), leading to a steeper mass increase with the semi-major axis. This is because our model tends to overestimate the collision rate of planets in these distant regions, where eccentricities are largely excited through frequent gravitational scatterings. Since our analytical formulation of secular evolution ignores higher-order terms in $e$ and $I$ in the disturbing function, the subsequent eccentricity evolution of such scattered eccentric planets may be less accurate.
\revise{Also, the eccentricity excitation during scattering events might be overestimated in the distant region.}
Nevertheless, the impact of this inaccuracy in distant regions on the evolution of 
inner ($a \lesssim$~1~au) planets is negligible.

In contrast to Class~I, the growth of Class II planets is minimal in both the NGPPS model and our model (panels (d) and (e)). These migrated planets are already massive in the initial state and relatively widely separated, as many are trapped in their mean motion resonances. Thus, Class II systems are more dynamically stable than Class I systems, a stability which has also been reproduced in our model.

In summary, the semi-analytical model successfully reproduces the mass and semi-major axis distribution of planetary systems shaped by giant collisions and gravitational scatterings across a wide range of orbital radii.

\subsection{Statistical features of the individual systems}


Figures~\ref{fig:all_stats_classI} and \ref{fig:all_stats_classII} compare the statistical features of planetary systems between our semi-analytical model and the NGPPS model for Class I and Class II systems, respectively. 
Here we have divided the planets of each system into four 
groups according to their semi-major axes, $<$~0.3~au, 0.3--1~au, 1--3~au, and 3--10~au, and compare the statistics in each region.
%
We find good agreement between the two models in all the metrics for both Class I and Class II systems, especially in inner regions of $\lesssim 3$~au.
This confirms that our model effectively predicts the statistics of the architectures of each planetary system, 
in addition to the overall distribution of mass and semi-major axis.

In the distant region ($\sim$3--10~au) of Class I systems, our model slightly underestimates the number of planets per system and overestimates the orbital separations and eccentricities.
\revise{
In this region, planetary growth is influenced by planetesimal accretion, which is not taken into account in our model. Moreover, as seen in Fig.~\ref{fig:Mp-a_comparison}, this population exhibits a large dispersion in planetary mass, requiring caution when comparing the results (we should note that the level of agreement of these statistics barely changes even when small planets with $<0.1M_\oplus$ are excluded). Nevertheless, some of the outcomes provide valuable insights into the limitations of our model.
For example,
}
the discrepancy \revise{in the number of planets} arises because our calculations show more ejection events than in the NGPPS results. While about 20\% of the Class I systems experience at least one ejection event in the NGPPS calculations, the fraction exceeds 60\% in our model, where most ejection events occur when a gravitationally scattered planet reaches $e>1$ during the subsequent secular evolution.
This discrepancy, \revise{along with the overestimation of final eccentricities, likely stems from an overestimation of eccentricity excitation following close-scattering events and limitations in the secular evolution model for highly eccentric orbits}, as noted in the previous section.
Further development of a model for the dynamical evolution of such eccentric planets is necessary to better predict the architecture of planets in distant regions, and also of planets around lower mass stars, where planets can easily become eccentric even in close-in regions.

\section{Discussion}
\label{sec:discussion}
\subsection{Effects of planetary radius}
\label{sec:planet_radius}

While our semi-analytical model can reproduce the results of the NGPPS models using direct $N$-body simulations for various planetary systems, its most critical assumption lies in the treatment of planetary radii. To assess the impact of planetary radii on dynamical evolution, we have performed calculations using a fixed bulk density of $5.5~{\rm g/cm^3}$, roughly corresponding to the case where envelope thickness is ignored. 

Panels~(c) and (f) of Fig.~\ref{fig:Mp-a_comparison} show the final distribution of planetary masses and semi-major axes in this case. We find that the treatment of planetary radii significantly alters the resultant distribution of planets. Specifically, when planetary radii are reduced (i.e., envelope thicknesses being ignored), more collisional events occur, 
resulting in a substantial decrease in the number of remaining planets compared to our fiducial case (panels (b) and (e)). Additionally, the typical planetary mass is also larger than the NGPPS results.
This is because smaller planetary radii lead to an increase in the escape eccentricity $e_{\rm esc}$ (see Eq.~\eqref{eq:e_esc}), which, in turn, 
enhances the excitation of orbital eccentricities during orbital crossing events.
Consequently, their semi-major axes change more drastically via gravitational scattering events, and the eccentricities after collisional events tend to be higher; both factors enhance the efficiency of collisional growth \citep[see also][]{Matsumoto+2021}.

How planetary envelopes affect the dynamical evolution of multiple-planet systems during the giant impact phase remains an open question and is beyond the scope of this study.
\revise{
Although the envelope of super-Earths and sub-Neptunes can significantly enlarge the planetary radii, the envelope mass and density is still small compared to those of the core~\citep[e.g.,][]{Lopez+Fortney2014}. 
}
This suggests that collision and accretion occur when the solid cores of the planets come into contact rather than their envelopes. 
Thus, collision rates can be overestimated by detecting collisions based on the planetary radii including the envelopes, which is also discussed in \cite{Emsenhuber+2021a}. 
To quantify the impact of planetary envelopes on dynamical evolution, we need to investigate the outcomes of collisions between enveloped planets using hydrodynamic simulations
\revise{\citep[e.g.,][]{Denman+2020}}. In particular, during the early stages following gas disk dispersal, planetary envelopes remain significantly inflated, affecting their dynamical evolution, while their influence becomes negligible in later stages, once the envelopes have contracted. 

\subsection{Effects of the stability of resonant planets}
\begin{figure}
    \centering
    \includegraphics[width=\linewidth]{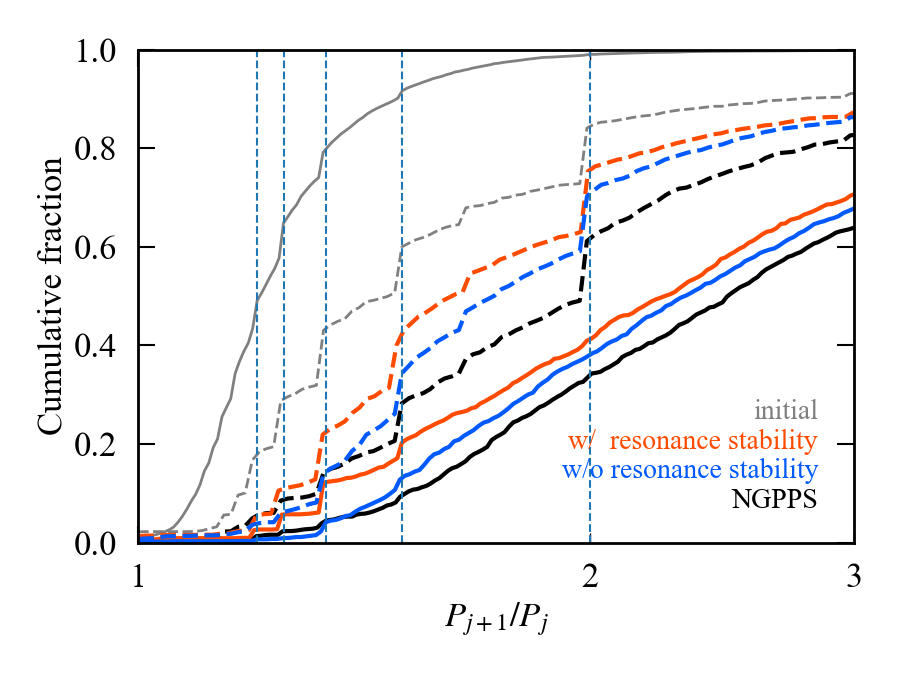}
    \caption{Cumulative number of planets normalized by the total number of planets as a function of the period ratio of neighboring planets at the initial state (gray), final state in our fiducial model (blue), in our model with resonance stability considered (red), and in the NGPPS results (black). The solid and dashed lines show results for Class I and II systems, respectively.
    Only planets within $\le 1$~au are considered.
    The vertical dashed lines indicate the locations of the 6:5, 5:4, 4:3, 3:2, and 2:1 resonances from left to right.}
    \label{fig:Pratio_cumhist}
\end{figure}

At the time of gas disk dispersal, many planets in the NGPPS data form resonance chains primarily composed of first-order resonances, as also suggested by other prevailing theories of planet formation \citep[e.g.,][]{Terquem+Papaloizou2007,Cresswell+Nelson2008,Cossou+2014}. 
The timing of 
the onset of orbital instability in such systems differs significantly from that in non-resonant systems. 
Specifically, chains with fewer than a critical number of planets are known to remain stable over the long term~\citep{Matsumoto+2012,Matsumoto+Ogihara2020,Pichierri+Morbidelli2020, Goldberg+2022}.

Here we show how the effects of resonance affect the final planetary distributions. 
To assess the stability of a resonant chain, we use the mass criterion for planets in the chain analytically derived 
by \cite{Goldberg+2022}:
\begin{equation}
    M_{\rm crit} = 0.2\qty(\frac{j-1}{j})^{1.5N_{\rm chain}} j^{1.2} (j-1)^{-6.2} M_*,
    \label{eq:Mcrit}
\end{equation}
where $j$ is the resonant index of the chain ($j:j-1$ resonance) and $N_{\rm chain}$ is the number of planets in the chain.
If the mass of the most massive planet 
in the chain is smaller than $M_{\rm crit}$, the chain is stable over the long term.
Here we only consider the first-order resonance.
When the chain consists of resonances with different indices $j$, the highest value of $j$ is used.
Then, when computing the time to the onset of orbital instability  
for three neighboring planets, $\tau_{\rm cross}$, we set $\tau_{\rm cross} = \infty$ if all three planets are in the resonance chain and the chain satisfies the above mass criterion.
Note that planets at the edge of the chain can cause orbital crossing with their neighboring non-resonant planets. 
For example, if planets 1 and 2 are in resonance and planets 2 and 3 are not, the orbital crossing timescale for this triplet is calculated as if all three planets were not in resonance.

We should also note that planets in mean motion resonance are identified solely on the basis of their orbital period ratios, without considering whether they are actually in resonance. In the circular restricted three-body problem, when a massless body $i$ is in a $j:j-1$ resonance with outer planet $i+1$, the resonance width in terms of orbital period is approximately expressed by~\citep{Murray+Dermott1999}
\begin{equation}
    \Delta P \approx (12\mu_{i+1} \alpha_{i} |f_{\rm d}(\alpha_i)|e_i)^{1/2} P_{i+1},
    \label{eq:deltaP_org}
\end{equation}
where $\mu_{i+1} = M_{\rm i+1}/M_*$, $\alpha_i= a_i/a_{i+1}$, and $f_{\rm d}$ is an interaction coefficient from the disturbing function. Since $f_{\rm d}$ is of order unity within the range of $\alpha_i$ we are interested in, we simplify Eq.~\eqref{eq:deltaP_org} to
\begin{equation}
    \Delta P \approx (10\mu_{i+1} \alpha_i e_i)^{1/2} P_{i+1}.
    \label{eq:deltaP}
\end{equation}
Although Eq.~\eqref{eq:deltaP_org} is not valid for low-eccentricity planets, we have confirmed that the approximation is sufficiently accurate for most planets in the initial NGPPS samples.
Thus, a neighboring planet pair is considered to be in resonance if the inner planet's orbital period is within the range of Eq.~\eqref{eq:deltaP} from any resonance positions of the outer planet.

\begin{figure*}
    \centering
    \includegraphics[width=\linewidth]{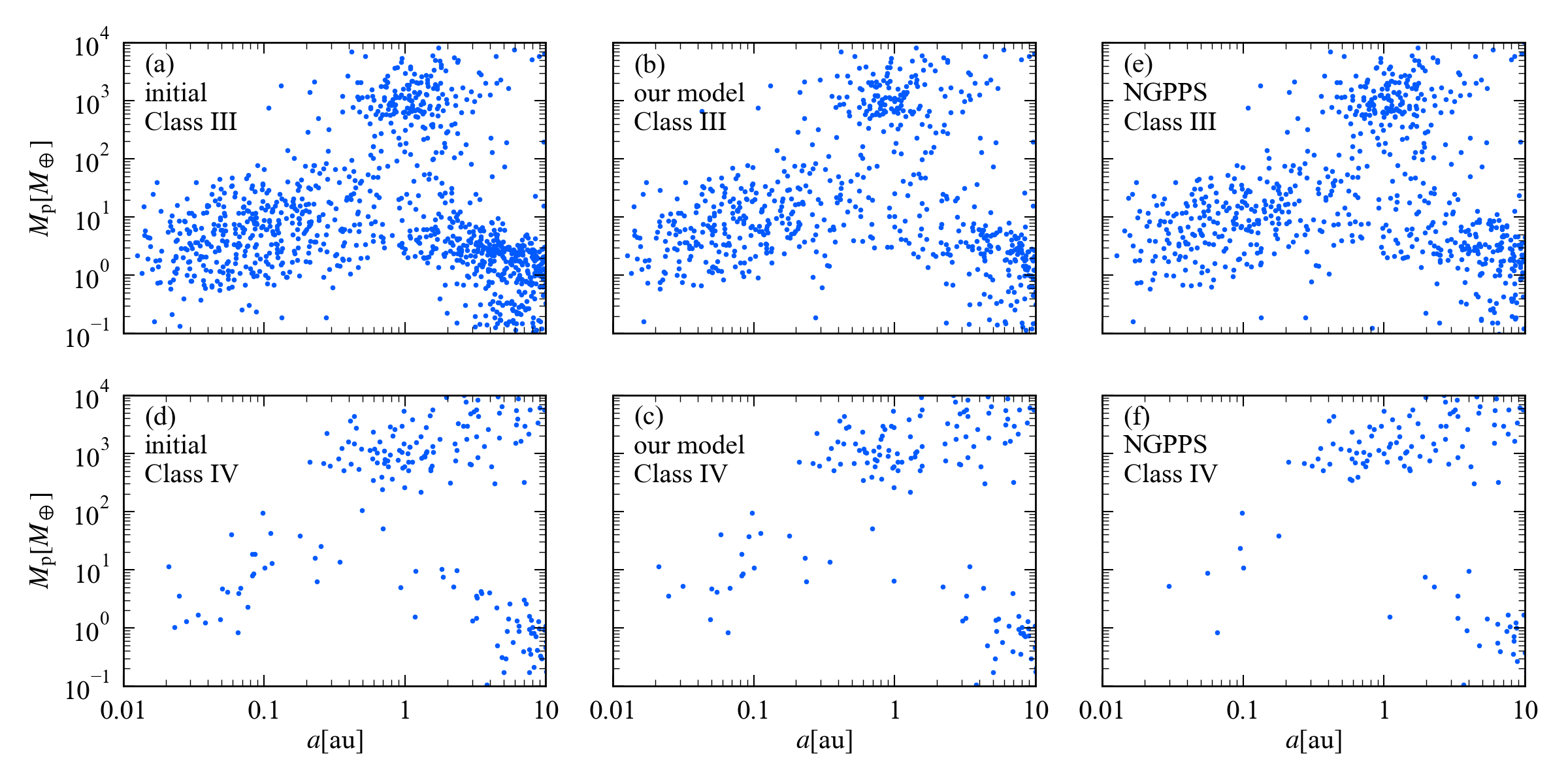}
    \caption{Mass and semi-major axis distribution of planets in Class~III (panels (a)--(c)) and Class~IV (panels (d)--(f)). 
    Each panel shows the distribution at the initial state ((a) and (d)), at the final state (100~Myr) in our model ((b) and (e)) and in the NGPPS model ((c) and (f)).
    }
    \label{fig:Mp-a_class34}
\end{figure*}

Figure~\ref{fig:Pratio_cumhist} compares the cumulative distribution of period ratios of neighboring planets in the Class I (solid lines) and Class II (dashed lines) systems. Results from our fiducial model (without considering the resonance chain stability) and the model including stability are shown, as well as the NGPPS results.
The distribution in the initial state is also plotted in gray for comparison. Here, we focus only on planets within 1~au.

In the initial state, many planets are closely packed and in resonance chains for both Class I and II. The NGPPS results show that most chains in Class I are destroyed via post-disk dynamical evolution, whereas many in Class II remain. 
Our fiducial model reproduces these features, and the overall distribution is quite similar, although the model slightly underestimates the abundance of planets in deep resonances, such as 5:4 and 4:3. 

However, the fraction of plants in resonance is overestimated when we include the stability criterion of resonance chains.
This is because the stability criterion of resonance chains in Eq.~\eqref{eq:Mcrit} accounts only for interactions between planets in the chain, while perturbations from 
other non-resonant planets can destabilize the chain. 
\revise{
In addition, when some planets in a chain cause an orbital crossing event, the remaining planets in the chain might not be in the resonance state anymore, depending on the details of the instability.  
}
Note that the overall mass and semi-major axis distribution of planets, as well as the statistics of individual systems, are hardly affected by the treatment of resonance chains, since the fraction of planets in resonance in the final state is small for both cases.

Thus, we have found that the model without the effects of resonance chain stability works better in our semi-analytical model.
Our results highlight the necessity of detailed investigations into the stability of planetary systems composed of resonance chains and non-resonant planets.

\begin{figure}
    \centering
    \includegraphics[width=\linewidth]{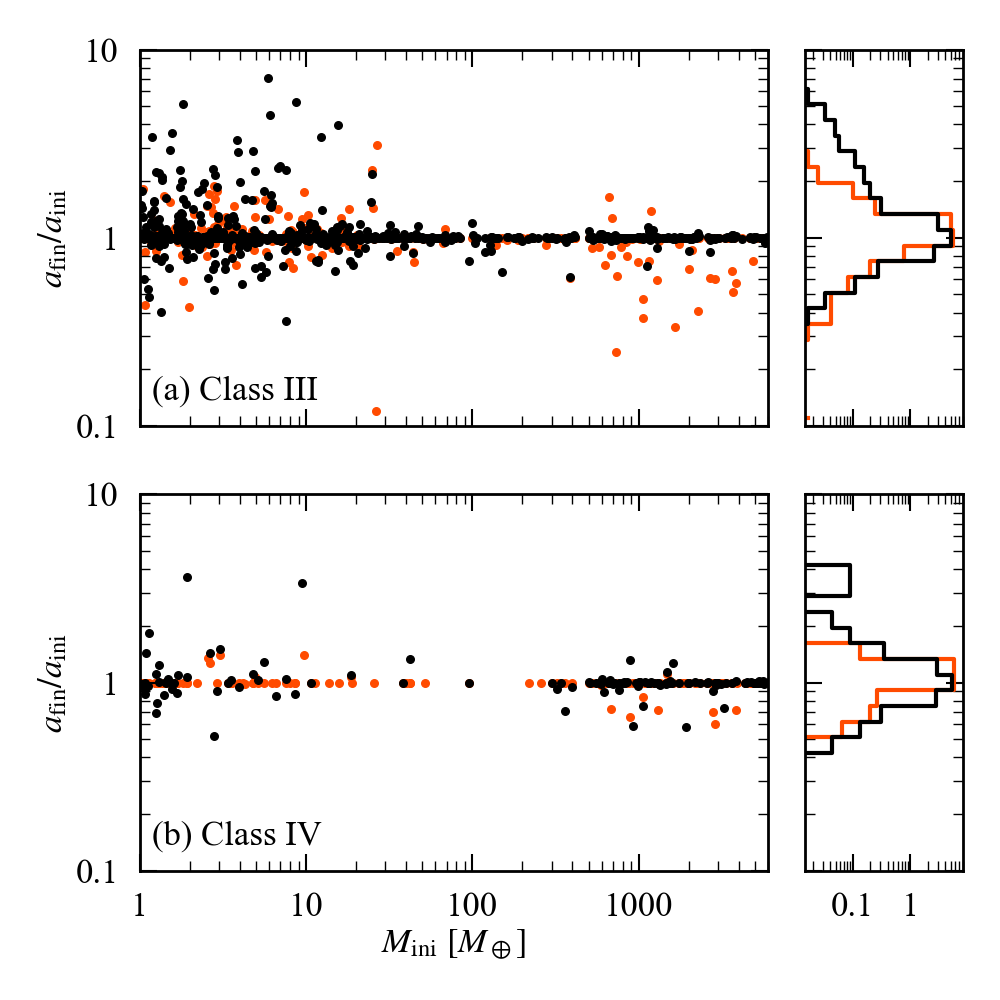}
    \caption{Ratio between the final and initial semi-major axes of planets in Class III (a) and Class IV (b) as a function of their initial mass. The \revise{orange} and \revise{black} points show results from our semi-analytical model and the NGPPS model, respectively. The right panels show the probability density distribution of $a_{\rm fin}/a_{\rm ini}$ in each class. 
    }
    \label{fig:a0aF_Mp}
\end{figure}
\subsection{Application to giant planet systems}

We investigated the applicability of our model to general planetary systems with low-mass planets. Here, we apply our model to systems with giant planets (Class III and IV) to test its robustness. 

Figure~\ref{fig:Mp-a_class34} shows the final mass and semi-major axis distribution of planets in Class III and IV systems in our model results and the NGPPS results, as well as the initial distributions.
The overall distribution is qualitatively similar, indicating that our model also works well for systems with giant planets.
However, the applicability is still not clear
because the number of Class III and IV systems is small (about 200 out of 1000 systems in the NGPPS results), and most of the systems are dynamically stable from our initial state (the moment of gas disk dispersal), as dynamical interactions already occurred earlier.

Therefore, we additionally compare the extent of changes in planetary systems from their initial state.
Figure~\ref{fig:a0aF_Mp} compares the ratio of the final $(a_{\rm fin})$ to initial $(a_{\rm ini})$ semi-major axis of planets in Class III and IV systems between our model and the NGPPS results. While many giant planets in the NGPPS results remain at their initial positions, our model overestimates the frequency of dynamical instability in giant planets, \revise{especially in Class III systems}. 
\revise{
Furthermore, we have found that the ejection of giant planets occurs significantly more frequently in our simulations than in the NGPPS results. In particular, nearly 20\% of Class III systems in our model experience at least one giant planet ejection, compared to only 8\% in the NGPPS simulations.
}

\revise{
These discrepancies fundamentally arise from our use of $e_{\rm esc}$ in describing dynamical evolution during close scattering events, which affects both the grouping method for computing $\tau_{\rm cross}$ and the post-scattering eccentricity estimation.
} 
Since the escape eccentricity $e_{\rm esc}$ of giant planets exceeds unity, most planets around giant planets are grouped together, reducing the orbital crossing timescale of the giant planet pair. In reality, surrounding small planets hardly affect the stability of giant planets, and two giant planets with a separation larger than $\sim 2\sqrt{3}r_{\rm H}$ are known to be stable~\citep{Gladman1993}. 
\revise{
Moreover, since we estimate the eccentricity after a scattering event using $e_{\rm esc}$, any scattering between two giant planets results in the ejection of one of them, leading to a significant overestimation of the ejection frequency.
}

\revise{
Thus, we need further improvements to apply our model to giant planet systems.}
Such improvements would also be important for planets around lower mass stars, where $e_{\rm esc} > 1$ is satisfied even for super-Earth-mass planets in close-in regions~\citep[e.g.,][]{Matsumoto+2020}.

\section{Conclusion}
\label{sec:conclusion}

We have improved the semi-analytical model for the dynamical evolution of low-mass planets to apply it to general planetary systems with larger dispersions in mass and orbital separation. The improvements include evaluating the number of planets that affect the orbital crossing timescale and ejection of planets. 

We validated our model by comparing it with the results of recent planet population synthesis calculations~\citep[][; referred to as NGPPS]{Emsenhuber+2021a,Emsenhuber+2021b,Emsenhuber+2023}, which use direct $N$-body simulations. Our semi-analytical model calculations used the same initial planetary distribution as the NGPPS calculations at the time of gas disk dispersal, focusing particularly on systems without giant planets.

Our model successfully reproduces the final distribution of the planetary mass and the semi-major axis obtained from the NGPPS. It also replicates various system architectures, such as the number of planets in a system, mean orbital separation and eccentricity, and intra-system dispersion of planetary mass. We also showed the applicability of our model to systems with giant planets.

Thus, we have confirmed that the semi-analytical model is applicable to various planetary systems derived from planet formation. Since we found that the planetary radius largely affects dynamical evolution, appropriate treatment of the collisional cross section, including the effects of primordial envelope, is important for predicting the outcome of post-gas disk dynamical evolution. The development of an integrated model that includes the formation and evolution of primordial envelopes, as well as other planet formation processes, will be addressed in future work.
Such a computationally efficient planet formation model enables conducting a large number of parameter studies and statistical comparison with observed exoplanet populations, which can contribute to advancing our understanding of planet formation.

\begin{acknowledgments}
This work is supported by JSPS KAKENHI Grant Nos.18H05439, 22KJ0816, 22K21344, 24K00698 and 24H00017. Part of the numerical computations were carried out on the general-purpose PC cluster at the Center for Computational Astrophysics, National Astronomical Observatory of Japan.
C.M. acknowledges the support from the Swiss National Science Foundation under grant 200021\_204847 “PlanetsInTime”. Part of this work has been carried out within the framework of the NCCR PlanetS supported by the Swiss National Science Foundation under grants 51NF40\_182901 and 51NF40\_205606. 
\end{acknowledgments}

\bibliography{refer}{}
\bibliographystyle{aasjournal}



\end{document}